\documentclass[conference]{IEEEtran}
\usepackage{graphicx}
\usepackage{subfigure}
\usepackage{graphics}
\usepackage{mathtools}
\usepackage{float}
\usepackage{ragged2e}
\usepackage{array}
\usepackage{tabulary}
\usepackage{hyperref}

\ifCLASSINFOpdf
\else
\fi

\begin{document}
%
\title{Emerging Phishing Trends and Effectiveness of the Anti-Phishing Landing Page}

\author{
    \IEEEauthorblockN{Srishti Gupta, Ponnurangam Kumaraguru}
    \IEEEauthorblockA{Indraprastha Institute of Information Technology, Delhi \\
    Cybersecurity Education and Research Centre (CERC), IIIT-Delhi
    \\\{srishtig, pk\}@iiitd.ac.in}
}

\maketitle
\begin{abstract}
Each month, more attacks are launched with the aim of making web users believe that they are communicating with a trusted entity which compels them to share their personal, financial information. Acquired sensitive information is then used for personal benefits, like, gain access to money of the individuals from whom the information was taken. Phishing costs Internet users billions of dollars every year. A recent report highlighted phishing loss of around \$448 million to organizations in April 2014. Researchers at Carnegie Mellon University (CMU) created an anti-phishing landing page supported by Anti-Phishing Working Group (APWG) with the aim to train users on how to prevent themselves from phishing attacks. It is used by financial institutions, phish site take down vendors, government organizations, and online merchants. When a potential victim clicks on a phishing link that has been taken down, he / she is redirected to the landing page. In this paper, we present the comparative analysis on two datasets that we obtained from APWG's landing page log files; one, from September 7, 2008 - November 11, 2009, and other from January 1, 2014 - April 30, 2014. We found that the landing page has been successful in training users against phishing. Forty six percent users clicked lesser number of phishing URLs from January 2014 to April 2014 which shows that training from the landing page helped users not to fall for phishing attacks. Our analysis shows that phishers have started to modify their techniques by creating more legitimate looking URLs and buying large number of domains to increase their activity. We observed that phishers are exploiting Internet Corporation for Assigned Names and Numbers (ICANN) accredited registrars to launch their attacks even after strict surveillance. We saw that phishers are trying to exploit free subdomain registration services to carry out attacks. In this paper, we also compared the phishing e-mails used by phishers to lure victims in 2008 and 2014. We found that the phishing e-mails have changed considerably over time. Phishers have adopted new techniques like sending promotional e-mails and emotionally targeting users in clicking phishing URLs.
\end{abstract}

\section{Introduction}
\begin{figure}[h]
\centering
\includegraphics[width=3.5in]{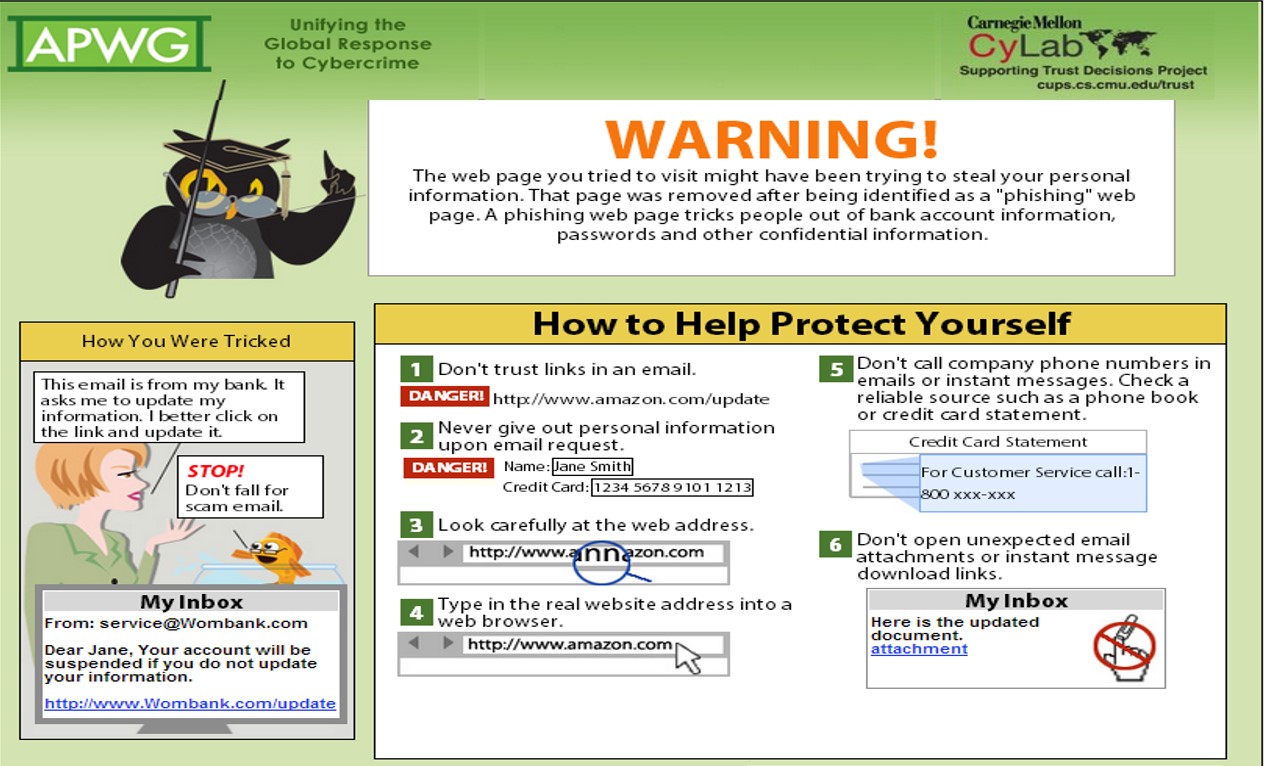}
\caption{APWG sponsored landing page developed at CMU in August 2008 to educate and train users not to fall for phishing attacks.}
\label{fig:education_page}
\end{figure}

With the increase in use of Internet for business, finance and personal investments, threats due to Internet frauds and eCrime are on rise. Internet frauds can take several forms, from stealing personal information to conducting fraudulent transactions. One interesting form of Internet fraud is phishing; Phishing is the act of attempting to acquire sensitive information such as usernames, passwords, and credit card details by masquerading as a trustworthy entity in an electronic communication \cite{jakobsson}. Phishing attacks use e-mail messages and websites designed to look as if they came from known and legitimate organizations, in order to deceive people in giving out their personal, financial or other sensitive information. A recent report by Anti-Phishing Working Group (APWG) showed that second half of 2013 saw a 60\% increase in phishing attacks from the first half \cite{apwg}. The report also highlighted an increase in the number of unique domain names and maliciously registered domain names for carrying out phishing attacks. This shows that criminals are exploiting best possible resources to carry out their tasks effectively. RSA, the security division at EMC\textsuperscript{2} recently released their monthly report on online fraud to show that phishing cost global organizations \$448 million in losses in April 2014. \footnote{http://www.ciol.com/ciol/resource-center/215305/phishing-cost-global-organizations-usd448-mn-losses-april-rsa} \newline\indent
Phishing has become a major concern for Internet Service Providers (ISPs), with pressure coming from both users who demand that service providers do more to protect them from attacks, and from the financial institutions targeted by these attacks. To reduce phishing damage, stakeholders have enacted their own countermeasures. ISPs, mail providers, browser vendors, registrars and law enforcement all play important roles. Although simple, due to off-the-shelf phishing kits provided by a thriving cybercrime ecosystem, phishing attacks can have a remarkable impact on the community. As a result, the anti-phishing solutions are increasing in number to reduce these attacks. APWG is an international consortium focused on eliminating fraud and identity theft that result from phishing, pharming and email spoofing of all types. APWG's Internet Policy Committee (IPC) worked with CyLab Usable Privacy and Security laboratory (CUPS) at the Carnegie Mellon University (CMU) in creation of APWG-sponsored anti-phishing landing page that aimed to educate users against the implications of phishing and guided them how to avoid it \cite{pk-404}.
The landing page, as shown in Figure \ref{fig:education_page}, was developed to train users who fell for phishing attacks. It was a substitute to a non-informative HTTP 404 (page not found) error page that does not provide any useful information to the end users. The landing page served as a repository of data that can be analysed to better understand phishing and its current trends.
\newline\indent
In this paper, we analysed two datasets from the landing page, first from September 2008 - November 2009, and second from January 2014 - April 2014 which were obtained from APWG, as Apache \footnote{http://www.apache.org/} redirect logs. This paper describes an evolutionary study on the two datasets, in two different timestamps to study phishing trends. Our objective in this research is to track the evolution of phishing and answer some questions; what new techniques have phishers incorporated to carry out their tasks? What are the characteristics of URLs and domains which are used to carry out these attacks, and what kind of registrars are being exploited by phishers to launch their attacks? What is the spread (across the globe) of the victims clicking on these phishing URLs? What are the places used by phishers in order to spread their phishing URLs?
\newline\indent
We received another dataset from APWG in the form of phishing e-mails reported by users to APWG. Through this dataset, we were able to determine the reasons that make them effective in convincing potential victims in giving their valuable information. We also measured the effectiveness of the landing page in helping users to avoid falling for phishing attacks. Our main contributions are:
\begin{itemize}
\item[$\bullet$ ] We present an evolutionary study on phishing techniques incorporated by phishers by comparing them from the September 2008 - November 2009 dataset and January 2014 - April 2014 dataset. We found that phishers have started to buy more number of domains, exploiting ICANN accredited registrars, and free subdomain registration services to launch attacks. We observed that phishing e-mails changed over time, with phishers using new techniques like propagation of promotional and money-related e-mails to con people.
\item[$\bullet$] We measured the effectiveness of educating page to see if they helped users not to fall for phishing. We found that 46\% users were benefited from the landing page as the number of phishing URLs they clicked in April 2014 reduced from January 2014.
\end{itemize}

The remainder of the paper is organized as follows: In Section 2, we discuss related research work. In Section 3, we present the infrastructure and framework used for our data collection from the landing page. In Section 4, we discuss the results from the landing page deployment. In the last section, we present the concluding remarks.

\section{Related Work}
Many researchers have examined the statistics of suspicious URLs to understand what leads to phishing.  Mc. Grath et al. performed a comparative analysis of phishing and non-phishing URLs \cite{mcgrath}. They studied features like IP addresses, WHOIS records, geographic information, and lexical features of the URL (length, character distribution, and presence of predefined brand names) and found different lengths for phishing and non-phishing URLs, misuse of free hosting services by phishers. Similar features were used by Guan et al. to classify URLs that appeared in Instant Messaging (IM) \cite{guan} Ma et al. built a URL classiﬁcation system that processed a live feed of labelled URLs and collected features (lexical, WHOIS features) for these URLs in real time \cite{ma} with an accuracy of 99\%. Zhang et al. built CANTINA, a tool which classified phishing URLs by analysing the content of the webpage \cite{zhang}. They assigned a weighted sum to 8 features (4 content-related, 3 lexical, and 1 WHOIS-related) to build the classifier. Among lexical features, they looked at dots in the URL, presence of certain characters, presence of IP address in the URL, and age of the domain. They further developed 8 discriminatory features and proposed CANTINA+ which explored HTML Document Object Model (DOM) and third party services to find phishing pages \cite{cantinaplus}. Miyamoto et al. used AdaBoost-based detection training sets to determine weights for the heuristics used in CANTINA and combined them using AdaBoost algorithm \cite{miyamoto}. Fu et al. tried to classify phishing web pages based on visual similarity \cite{yu}. They compared potential phishing pages against actual pages and assessed visual similarities between them in terms of key regions, page layouts, and overall styles.
\newline\indent
Blum et al. proposed a method to detect phishing URLs based on SVM \cite{blum}. They used 23 features to train the SVM based on protocol, domain, and path features of the URL. They achieved an accuracy of 99\%. Fette et al. used machine learning to classify phishing messages \cite{fette}. They used the properties of URLs present in the message (e.g., the number of URLs, number of domains, and number of dots in a URL) and could identify suspicious URLs with 96\% accuracy. Bergholz et al. further improved the accuracy of Fette et al. by introducing models of text classiﬁcation to analyse e-mail content \cite{bergholz}. They trained the e-mail features using Dynamic Markov Chains and Class - Topic models. Whittaker et al. analysed URL and contents of the page to determine whether a page is phishing or not \cite{colin}. They used features like presence of IP address, string characteristics of the URL and could classify more than 90\% phishing pages. Kolari et al. used URLs found within a blog page as features to determine whether the page is spam with good accuracy \cite{kolari}. We used some URL based features (like number of dots, number of subdomains in the URL, presence of IP address) and certain domain related features (information obtained from WHOIS records) to find the change in URL structure and domain characteristics in our datasets.
\newline\indent
Highlighting the need for anti-phishing solutions, researchers first tried to understand why phishing works, its economic and psychological impact. Dhamija et al. gave a psychology based discussion on why people fall for phishing \cite{dhamija}. They analysed 200 phishing attacks and identified several reasons, ranging from pure lack of computer system knowledge, to visual deception tricks used by adversaries, due to which users fall for phishing attacks. They conducted a usability study to show that people generally don't look at the browser based cues like address bar, security indicators etc. Downs et al. explored the mental models used by the Internet users to evaluate potential phishing pages \cite{downs}. Some of their subjects used incorrect strategies to analyse potential scams, leaving them at risk. Fogg et al. studied the attributes of web pages that make it credible \cite{fogg}. They found that many features of a page's appearance enhance its perceived credibility, a fact that phishers routinely exploit. Moore et al. gave an economic model to characterize the trade-offs between advertising and malware as monetary vectors providing insights into the economic impact of phishing attacks \cite{moore}. Al-Momani et al. developed a model to classify e-mails into phishing e-mails and legitimate e-mails in online mode \cite{momani}. Spamassassin was built with a number of rules to detect features common in spam e-mail that go beyond the text of the email. \footnote{http://spamassassin.apache.org/} Such text included things like the ratio of pixels occupied by text to those occupied by images in a rendered version of the e-mail, presence of certain fake headers, and the like. In our work, we study the features in phishing e-mails which convinced people to respond and click them.
\newline \indent
Anti-phishing is the countermeasure to defeat phishing. There are a number of countermeasures proposed to combat phishing. Kirda et al. developed a browser extension AntiPhish, that aimed to protect users against spoofed website-based attacks \cite{kirda}. Several other toolbars like SpoofGaurd \cite{chou}, TrustBar  \cite{herzberg}, PhishZoo \cite{afroz}, Netcraft \cite{netcraft}, and SiteAdvisor \cite{siteadvisor} were developed to warn users about phishing attacks. Dhamija et al. developed ``trusted paths" for the Mozilla web browser that were designed to assist users in verifying that their browser has made a secure connection to a trusted site \cite{dhamijatygar}. In our work, we see the effectiveness of browser blacklists in preventing victims not to click phishing links.
\newline\indent
 Another approach has been to educate and train users about phishing. Kumaraguru et al. used online training materials to teach people how to protect themselves from phishing attacks \cite{pk-404}. Robila et al. educated users using phishing IQ tests and class discussions \cite{robila}. They displayed legitimate and fraudulent e-mails to users and had them identify the phishing attempts from authentic e-mails. It helped users in knowing what to look in the e-mails. Jagatic et al. developed a contextual training approach in which users sent phishing e-mails to probe their vulnerability \cite{jagatic}. At the end of the study, users were typically given additional materials informing them about phishing attacks in general. This approach had been used at Indiana University in studies conducted on students about contextual attacks making use of personal information. We measure the effectiveness of real-time training provided by the landing page to help users prevent clicking phishing URLs.
\newline\indent
The network characteristics of spam has been investigated by spammers. Anderson et al. focussed on the Internet infrastructure use to host phishing scams \cite{anderson}. They found that large number of hosts are used to advertise Internet scams using spam campaigns, individual scams themselves are typically hosted on only one machine. Ramachandran et al. studied the network level behaviour of spammers, like IP address ranges that send out the most spam, common spamming modes, persistence of spamming hosts, and botnet spamming characteristics \cite{ramachandran}. Casado et al. used passive measurements of packet traces captured from spam sources to estimate the bottleneck bandwidths of TCP flows from these spam sources \cite{casado}. Jung et al. studied the DNS blacklist traffic to monitor the IP addresses that were sending out spam. They also observed the activity distribution of spam source hosts \cite{jung}. In our work, we study the machines (hosting infrastructure) that are used to host phishing campaigns.
\newline
\indent
Most of the past work has been done on phishing detection, building classifiers to predict phishing URLs and e-mails, economic impact, phishing trends, network characteristics, anti-phishing solutions and psychological aspects of phishing. We built our analysis on the work done by Kumaraguru et al. \cite{pk-404} and recently generated APWG report \cite{apwg} to do a longitudinal study on true positive (close to ground datasets) to understand the evolution of techniques used by phishers in order to spread phishing URLs. We used two APWG's datasets obtained from phishing landing page, one from September 2008 - November 2009, and other from January 2014 - April 2014, to draw our comparative analysis. We also study the features of phishing e-mails that compel users to click on phishing links. In this paper, we also look at the effectiveness of landing page to help users prevent themselves not to fall for phishing attacks.

\section{Infrastructure}
APWG and Carnegie Mellon CyLab's Supporting Trust Decisions Project launched the Phishing Education Landing Page Program \footnote{http://phish-education.apwg.org/r/about.html} in August 2008 to educate users about phishing. The goal of this initiative was to instruct consumers on online safety at the ``most teachable moment" i.e. when they have just clicked on a link in a phishing communication \cite{pk-phd}. This landing page was hosted on APWG servers and is now translated into more than 20 languages. Users would be redirected to the specific language version of the page depending on the default language of their web browser. The ISPs, registrars or any other organization who had control of handling phishing pages were asked to redirect their user base to the landing page. When an ISP shuts down a phishing website / page and a user clicks on the link to that page, he / she gets to see a ``HTTP 404 not found page". For ISPs who opted for this initiative, their users were redirected to the landing page instead of `not found' page. To ensure proper redirection by the take down vendors, a set of instructions were furnished in a ``how to" file which was available on the website page.~\footnote{http://phish-education.apwg.org/r/how\_to.html} While doing the redirect, the ISPs were suggested to add the phishing URL as a parameter in the URL requesting the landing page. This was done by putting the phishing URL after ``?" in the HTTP request to the landing page.
\newline
\indent The APWG's server access log records all the requests in Apache's log format. The records were logged in the following Apache log format: ``\textbraceleft \%h \%l \%u \%t \textbackslash``\%r\textbackslash" \%s \%b \textbackslash``\%\textbraceleft Referrer\textbraceright i\textbackslash"~\textbackslash``\%\textbraceleft User-agent\textbraceright i\textbackslash"\textbraceright" i.e., each request was logged with the originating IP address, date, requesting URL, success code, size of the request header, and browser / client information. By mining the landing page log files, we could create a list of phishing URLs that were redirected to the landing page. We computed metrics based from these logs and report results for time period January 1, 2014 - April 30, 2014. In rest of the paper, we will refer this as \textit{2014 dataset}. We also analysed redirect logs on the landing page from September 7, 2008 - November 11, 2009 to study the evolution of techniques used by phishers in order to attract maximum victims to the phishing URLs. We will refer this as \textit{2008 dataset} in rest of the paper. We correlated the data from log files to the phishing e-mail feed archive by APWG to find out which e-mails led the users to visit the landing page. We studied how successful the landing page was to help users not to fall for phishing attacks.
\newline\indent
The data collected from the log files does not represent the entire population of the people clicking on phishing URLs and e-mails. If a user clicks on a phishing link and the link is already in the browser blacklists, then he / she will be blocked by browser settings and will not be redirected to the landing page. Some take town vendors may stop redirecting users to the landing page and show them the original `HTTP 404 not found page'. Thus, our data is a good lower bound for people who clicked on the phishing URLs and e-mails.

\section{Analysis}
In this section, we present the detailed analysis that we performed to answer our research questions. We discuss the structure of phishing URLs, insights into the domain used for creating URLs, geographical spread of victims clicking the phishing URLs and countries hosting the phishing domains, learning curve of the users after getting trained from the landing page, distribution of user agents like Internet Explorer, Firefox etc. used by the victims while clicking the phishing URLs, and referrer analysis to get information about the places used for spreading these links. We also present feature analysis performed on the e-mails to study the change in phishing e-mails from 2008 to 2014.

\subsection{Statistics on Apache logs}
To analyse results from the logs, we used only the entries that contained ``/r/'', because the entries were created in the logs when users clicked on links to websites that were taken down. The entries having ``?'' in the HTTP request were considered to include the phishing URL; the vendors were asked to add user-clicked URL (phishing URL) after a ``?'' in the HTTP request which they sent to the landing page. The detailed architecture of how the landing page works in given in Figure 1 \cite{pk-404}. We removed the entries that contained terms like `ORIGINAL\_PHISH\_URL' or `www.phishsite.com' or `the-phishing-page.html.' These were used in the documentation on how to implement the landing page and were likely to be hits from people testing the landing page. To ensure that clicks to the landing page were received by the end users and not take down vendors, we considered the URLs having greater than 5 hits in \textit{2014 dataset}. After analyzing the frequency distribution of the hits corresponding to each URL, we found a significant drop in URLs with 5 hits compared to the ones getting more than 5 hits. We assume that URLs which appeared in the logs less than five times were mostly used by take down vendors or organizations testing their implementation of the landing page or checking whether the landing page is still active. Table \ref{tab:url stats} presents statistics for the total hits received on the landing page. There were 3,613,410 total hits on 10,833 unique URLs redirected to the landing page. For the \textit{2008 dataset}, we considered URLs having hits greater than 5 after looking at the frequency distribution curve and recorded 2,977,052 total hits on 21,890 unique URLs. By this, we have kept the hits in both the analysis to be same, greater than 5.

\begin{table}[h]
\caption{Comprehensive view of the APWG landing page logs for the period January 1, 2014 to April 30, 2014.} \label{tab:url stats}
\begin{small}
\centering
\begin{tabular}{|>{\raggedright}p{3cm}|>{\raggedright}p{1.15cm}|>{\raggedright}p{1.75cm}|p{1.5cm}|} \hline
\small Statistics &\small Whole dataset &\small \textless= 5 hits &\small \textgreater  5 hits \\ \hline
Number of unique URLs & 28,471 & 17,638 & 10,833  \\ \hline
Total Hits for all unique URLs & 3,646,483 & 33,073 & 3,613,410 \\ \hline
Maximum number of hits for a single URL & 342,317 & 5 & 342,317 \\ \hline
Minimum number of hits for a single URL & 1 & 1 & 6 \\ \hline
Average number of hits per URL & 104.9	& 1.6 & 300.2 \\ \hline
Median number of hits per URL & 2 & 1 & 17\\ \hline
Standard Deviation for the URLs & 3077.2 & 1.1 & 5224.5 \\ \hline
\end{tabular}
\end{small}
\end{table}

\subsection{Location Analysis}

\begin{figure}[h]
\centering
\includegraphics[width=3.5 in]{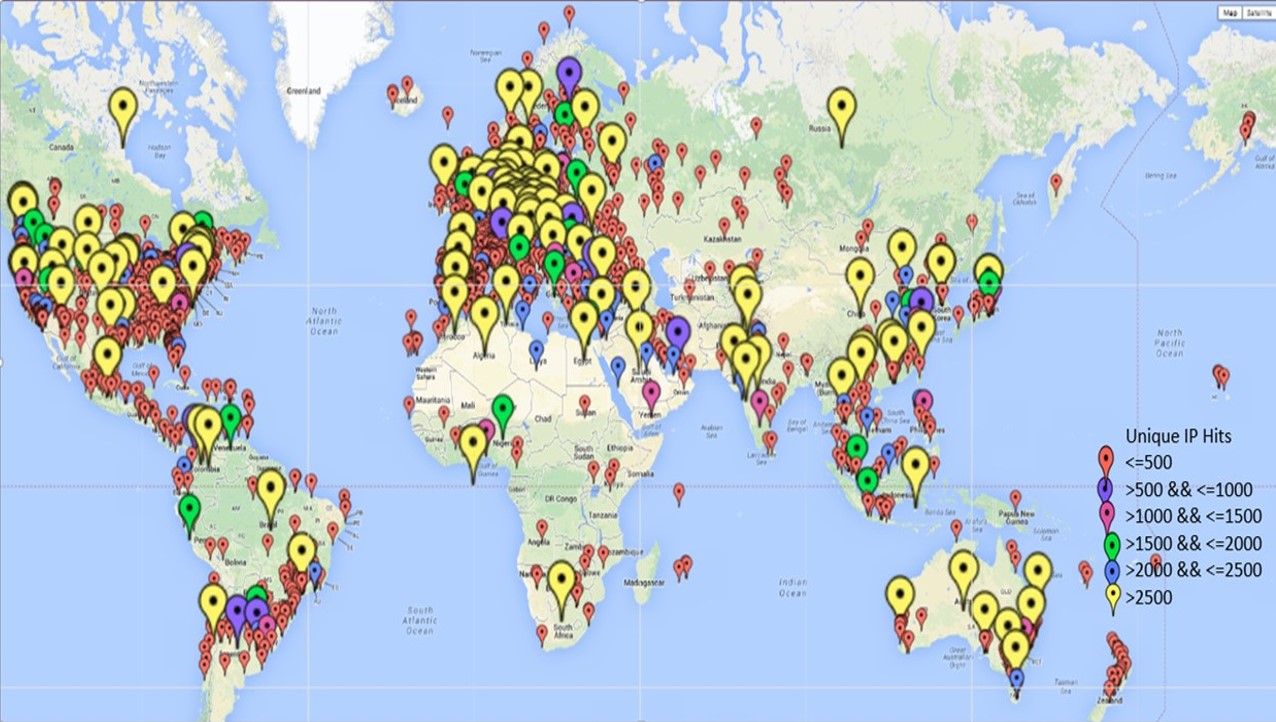}
\caption{Location information for the clicks obtained in \textit{2014 dataset}, shows that the United States is most vulnerable to phishing attacks, followed by France and Germany.}
\label{fig:geomap}
\end{figure}

\begin{figure}[h]
\centering
\includegraphics[width=3.5in]{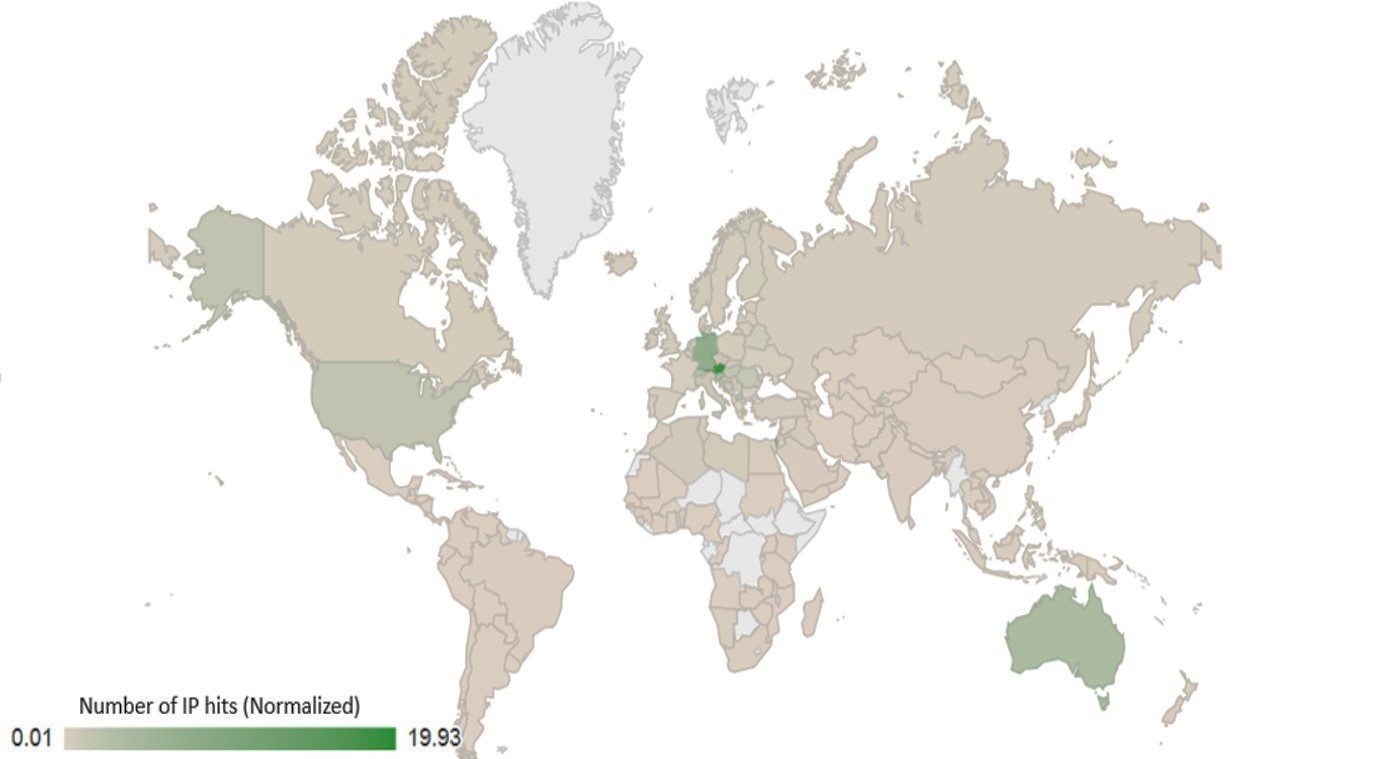}
\caption{Location information for the clicks obtained in \textit{2014 dataset}, normalized for different Internet population; shows that Austrians have been the most gullible followed by French and Germans.}
\label{fig:normalized_hits}
\end{figure}

To find the geographical reach of phishing URLs across the globe and get an idea about the countries being infected / vulnerable to phishing attacks, we fetched the latitude and longitude information corresponding to each IP address observed in our dataset. We used a public REST API, \footnote{http://freegeoip.net/} to get the geo-location (latitude, longitude) information corresponding to each unique IP address. Figure \ref{fig:geomap} shows the distribution of hits from the countries across the globe on world map. The United States, France, Germany, Australia were found to be most vulnerable to phishing attacks. However, the United States has a large Internet penetration rate, so to find the percentage of users who were vulnerable to phishing attacks in each of these countries, we divided the number of clicks with its respective Internet population. \footnote{http://www.census.gov/population/international/data/idb/rank.php} Figure \ref{fig:normalized_hits} shows the normalized distribution of clicks for 168 unique countries found in the \textit{2014 dataset}. We found that countries like Austria, France, Europe had higher percentage of Internet users falling for phishing attacks. In \textit{2008 dataset}, we received clicks from 167 unique countries where the top 5 maximum hits were received from Peru, the United States, Venezuela, Argentina, and Japan.
\newline\indent
Over recent years, there has been greater emphasis on using local language of the target audience. For example, phishers construct their e-mails using the appropriate language for the target audience. \footnote{http://www-935.ibm.com/services/us/iss/pdf/phishing-guide-wp.pdf} To see if phishers incorporated similar technique, we studied if there exists any correlation between the location from where user clicked the phishing URL and the language in which the landing page was displayed. For example, the page was requested in Arabic for people who belonged to countries like Iraq, Saudi Arabia, Kuwait, United Arab Emirates, and the like. We looked for the countries and the languages in which they were redirected, and compared each country with the list of its official languages as provided on Wikipedia. \footnote{http://en.wikipedia.org/wiki/List\_of\_official\_languages} We searched the requested URL for two character language code (e.g., `en' for English) to find out the language in which the landing page was requested. We found 11 unique languages in the dataset; Arabic, Dutch, English, French, German, Hungarian, Italian, Japanese, Norwegian, Russian, and Spanish. We found that all languages except English had users from those parts of the world where the particular language was the official language. For English, however, we received clicks from nearly all parts of the world.

\subsection{URL Analysis}
\begin{figure}[h]
\centering
\includegraphics[width=3.5in]{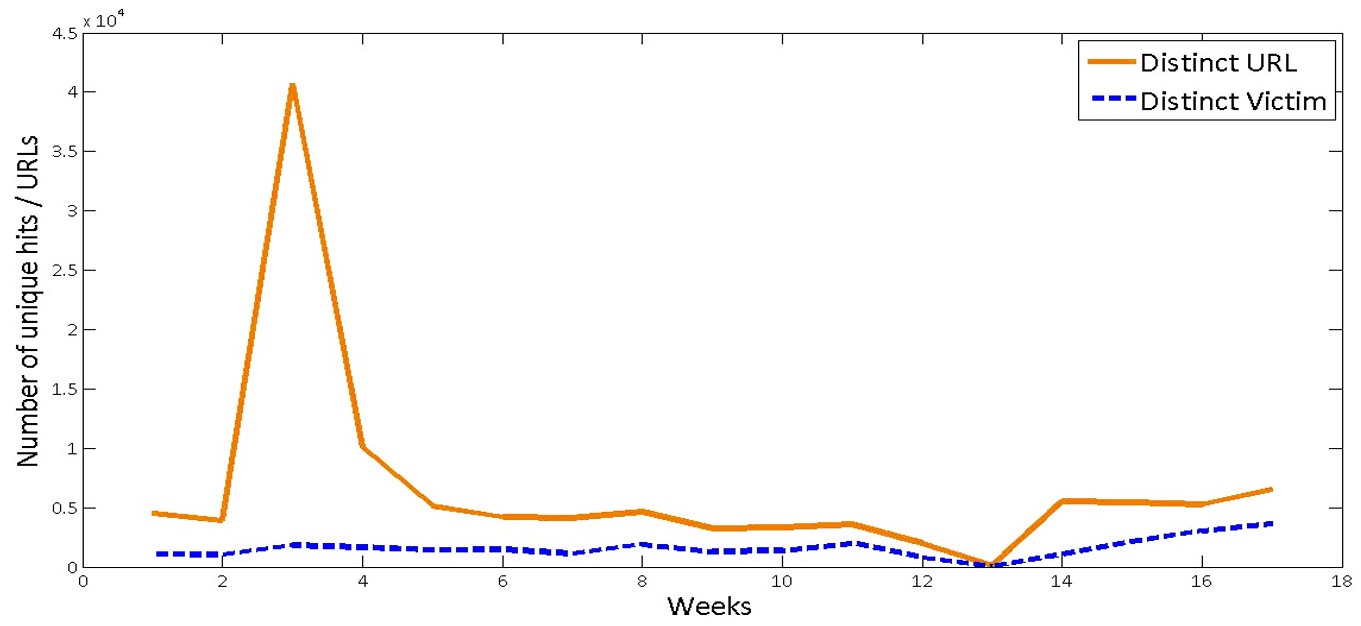}
\caption{Analysis of phishing campaigns during January 2014 - April 2014, by plotting the number of distinct URLs and IP hits per week (17 weeks). Phishing campaigns were found to be always successful since the number of URLs always exceeded the number of victims falling for it.}
\label{fig:phishing_campaign}
\end{figure}

\begin{figure*}[t]
\centering
\subfigure[Number of clicks less than 500.]
{
\label{fig:500}
\includegraphics[width=5cm,height=7cm,keepaspectratio]{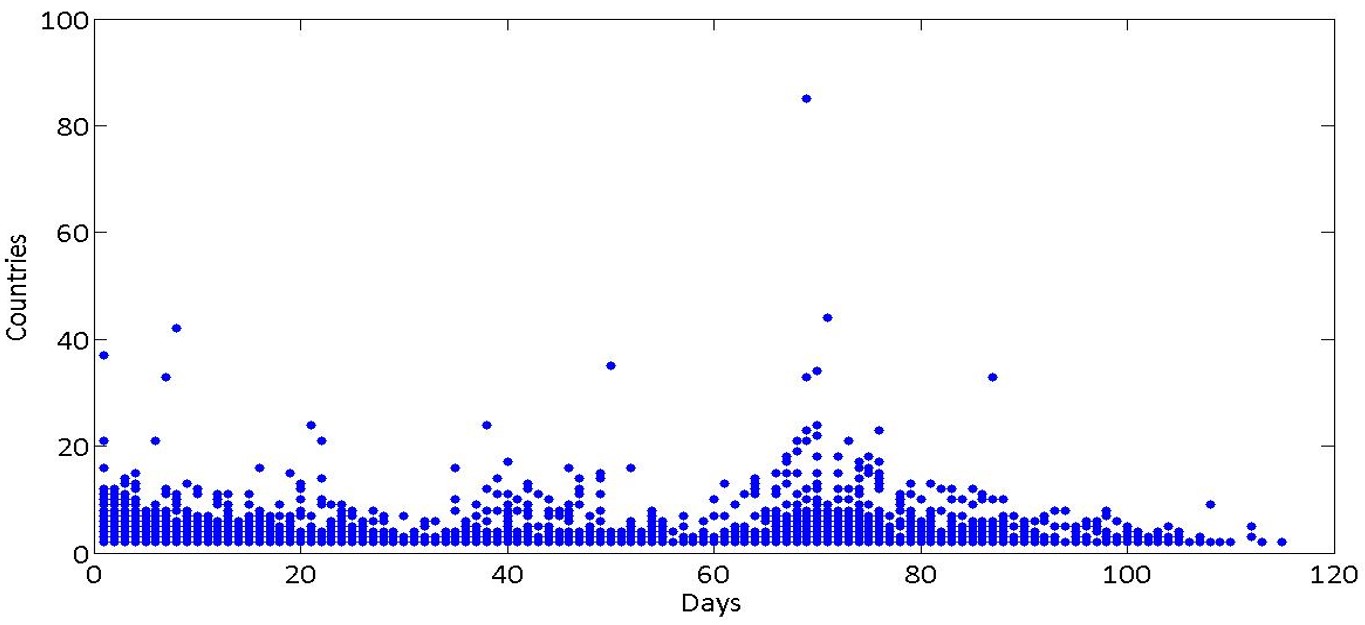}
}
\subfigure[Number of clicks between 500 and 1500.]
{
\includegraphics[width=5cm,height=5cm,keepaspectratio]{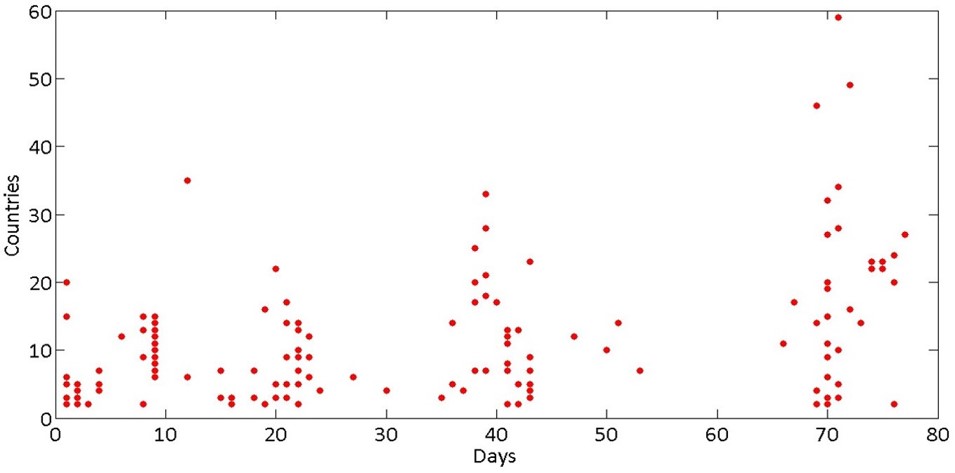}
\label{fig:1500}
}
\hfil\hfil
\subfigure[Number of clicks between 1500 and 5,000.]
{
\includegraphics[width=6.5cm,height=5.5cm,keepaspectratio]{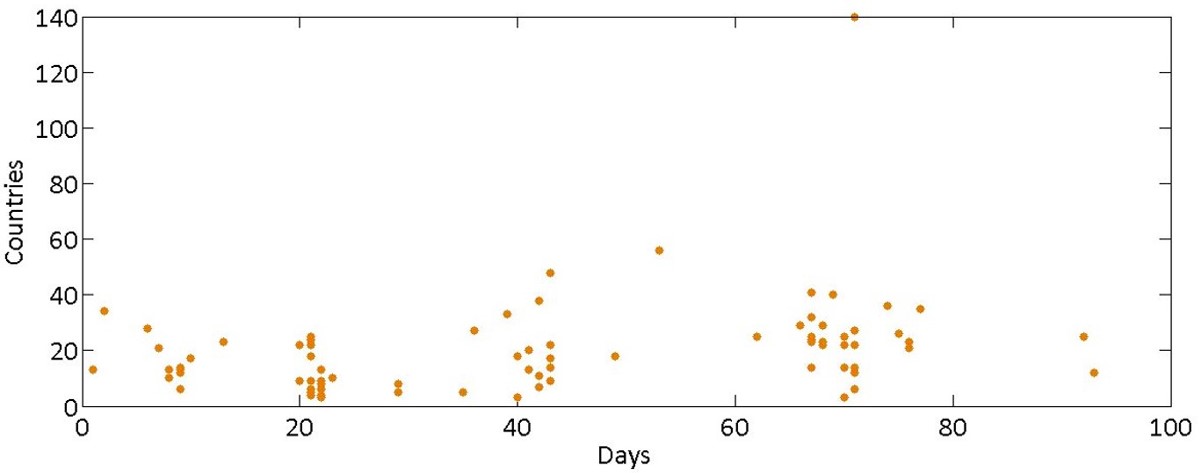}
\label{fig:5000}
}
\hfil\hfil
\subfigure[Number of clicks between 5,000 and 10,000.]
{
\includegraphics[width=6.5cm,height=5.5cm,keepaspectratio]{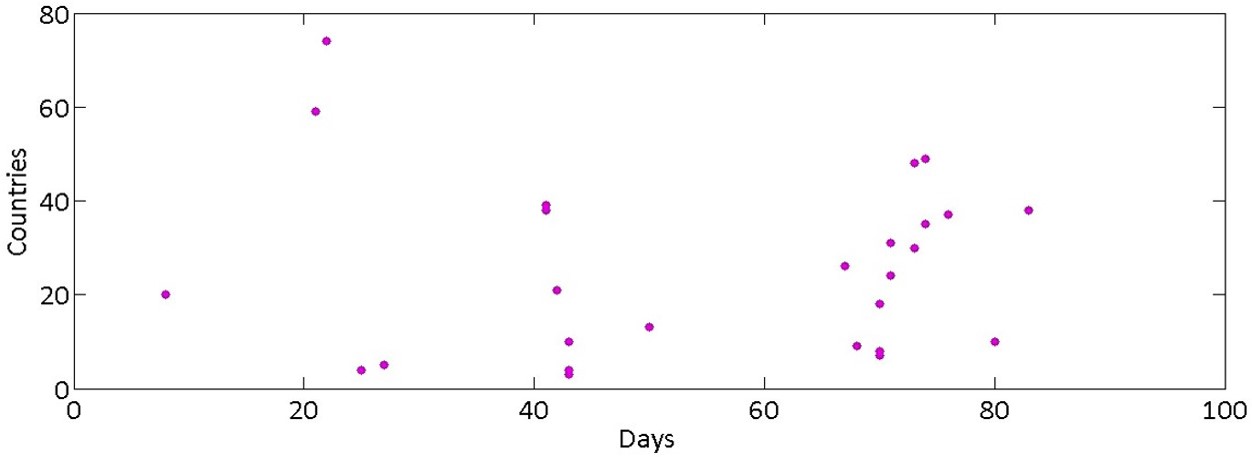}
\label{fig:10000}
}
\hfil\hfil
\subfigure[Number of clicks greater than 10,000.]
{
\includegraphics[width=6.5cm,height=5.5cm,keepaspectratio]{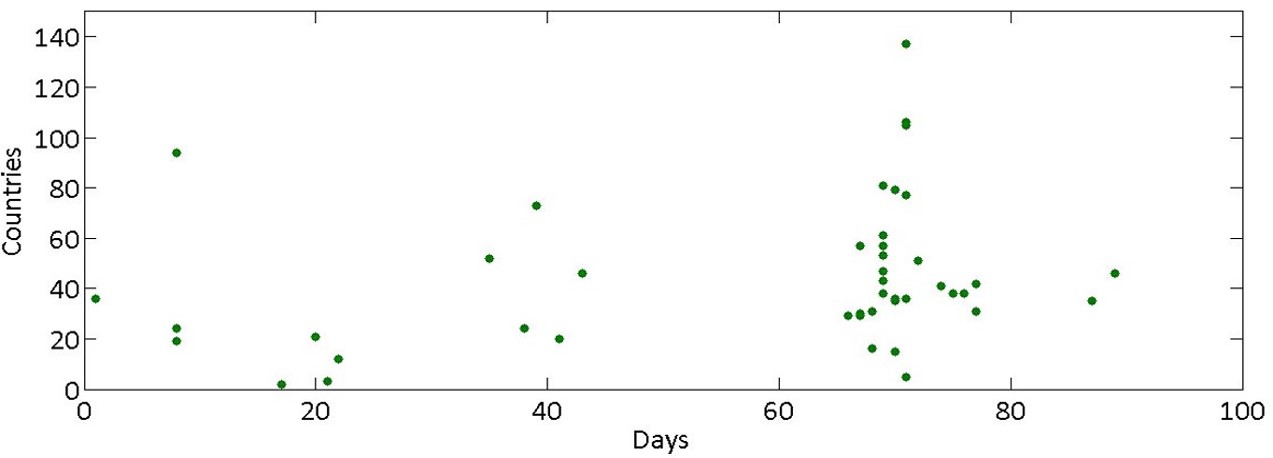}
\label{fig:g10000}
}
\hfil\hfil
\caption{Scatter plot of phishing URL for geographical and temporal spread as observed in \textit{2014 dataset}. It shows that URLs having moderate number of clicks spread evenly, geographically and temporally. URLs which receive lowest clicks remain active for longer duration and spread across lesser countries, while URLs with highest clicks span greater number of countries and have average lifetime.}
\label{fig:scatter_plot}
\end{figure*}

In a URL based phishing attack, an adversary lures the victim into clicking a URL pointing to the phishing site. Typical phishing URLs have the following structure,~\textit{http://domain.TLD/directory/filename?query-string}. The directory specifies the path with the file which is passed with a query string, together forming the pathname of the URL. The adversary usually obfuscates the URLs using several methods. We discuss some of the techniques currently in use.
\newline\indent
(i) \textit{IP address obfuscation}:  For a web browser to communicate over the Internet, the domain address must be resolved to an IP address. This resolution of IP address to host name is achieved through domain name servers. A phisher may wish to use the IP address as part of a URL to obfuscate the host and possibly bypass content filtering systems, or hide the destination from the end user. \footnote{http://www-935.ibm.com/services/us/iss/pdf/phishing-guide-wp.pdf} Garera et al. showed that the blacklist URLs generally have a significant percentage of IP address in the URL pathname \cite{sujata}. We used regular expression to check the presence of an IP address in a phishing URL in our dataset. We did not find significant difference in presence of such URLs in \textit{2008} and \textit{2014 datasets}. This shows that the number of attacks using IPs remained steady. As phishing attacks are becoming more sophisticated, IP-based links are becoming less prevalent, and attackers are buying domain names to make their URL look more genuine and legitimate.
\newline\indent
(ii) \textit{Directory structure similarity}: As pointed out by Prakash et al., there is a good chance that phishing URLs share a common directory structure, but with different filename or query string \cite{phishnet}. This helps phishers to launch different kind of phishing URLs keeping the directory structure same. We maintained a path equivalence class in which URLs with similar directory structure were grouped together. The URL was parsed for each `/' and string matching algorithm was used to compare the directories. We found that 38\% of phishing URLs in the \textit{2014 dataset} had the same directory structure with either different domains or different query string in the end to add variations. This is higher than 18\% observed in our \textit{2008 dataset}. This could be possible if phishers use single machine to launch multiple URLs or share common public directory in a network to carry out attacks.
\newline\indent
(iii) \textit{Number of host components in phishing URL}: Phishers normally use a long hostname, to confuse viewers into believing that their webpage is legitimate. They try to append an authentic-sounding word to their domain name. For eg., appleid.apple.co.uk.cgi-bin.webobjects.myappleid.woa.verify.appleid-serv.co.uk. This is easy to detect automatically by counting the number of host segments before the domain (9 in this case). Whittaker et al. showed that URLs where the number of host components is greater than 3 are generally phishing pages \cite{colin}. Our \textit{2014 dataset} had 17.4\%   URLs having length greater than 3 compared to 7.8\% observed in \textit{2008 dataset}. This shows that phishers have started using more sophisticated techniques than that were used earlier to compel users in clicking phishing URLs.
\newline
\newline\indent
We analysed how phishers propagated their phishing campaigns by dividing our \textit{2014 dataset} into 17 weeks. We plotted total number of unique URLs and unique IP hits observed per week. We assumed that each unique IP address corresponds to a unique user. Our results would have been more accurate if we would have obtained information about the users from the take down vendors. As shown in Figure \ref{fig:phishing_campaign}, the number of IP hits (victims) were always found to be greater than the number of URLs which indicates that phishers were successful in luring their victims. To analyse the peak observed in the third week, we calculated the ratio of URLs to victims for each day in that particular week. We found that more than 90\% victims were captured on January 19, 2014. Next, we looked at the geographical distribution of these victims; majority were found from Germany, Austria and the United States (\textit{the most vulnerable countries as discussed in the previous section}). We further looked at the top 10 URLs which received maximum clicks in our \textit{2014 dataset}; we observed that these URLs had highest number of victims from Germany and followed a particular format i.e., \textit{string1.string2.com/telekom} and \textit{string1/vodafone\_online/}. This outrage was a result of the malicious campaign which included e-mails that masqueraded as bills from NTTCable and from VolksbankU. \footnote{http://blogs.cisco.com/security/fake-phone-bills-contain-malware-targeting-dt-customers/} The malware that started on January 5, 2014, recorded the largest number of attacks in Germany. We observed a drop in thirteenth week due to some technical issues with the APWG servers, as a result of which the redirect requests could not be logged for that week.
\newline\indent
To check correlation between a URL's geographical spread and its lifetime (days between first and last click during the period 01/ 01/ 2014 - 30/ 04/ 2014), we plotted Figure \ref{fig:scatter_plot}. Each point is a URL whose x and y is determined by its lifetime and geographical spread respectively. Figure \ref{fig:500} showed that large percentage of URLs received lowest number of clicks, spanned across lesser number of countries, and remained active for a longer duration. Figure \ref{fig:g10000} showed that few URLs that received highest clicks, spread across larger number of countries, and had an average lifetime. URLs receiving intermediate number of clicks spread evenly, geographically and temporally as shown in Figure \ref{fig:1500}, \ref{fig:5000}, and \ref{fig:10000}.

\begin{figure*}[t]
\centering
\subfigure[]
{
\includegraphics[width=3.5in]{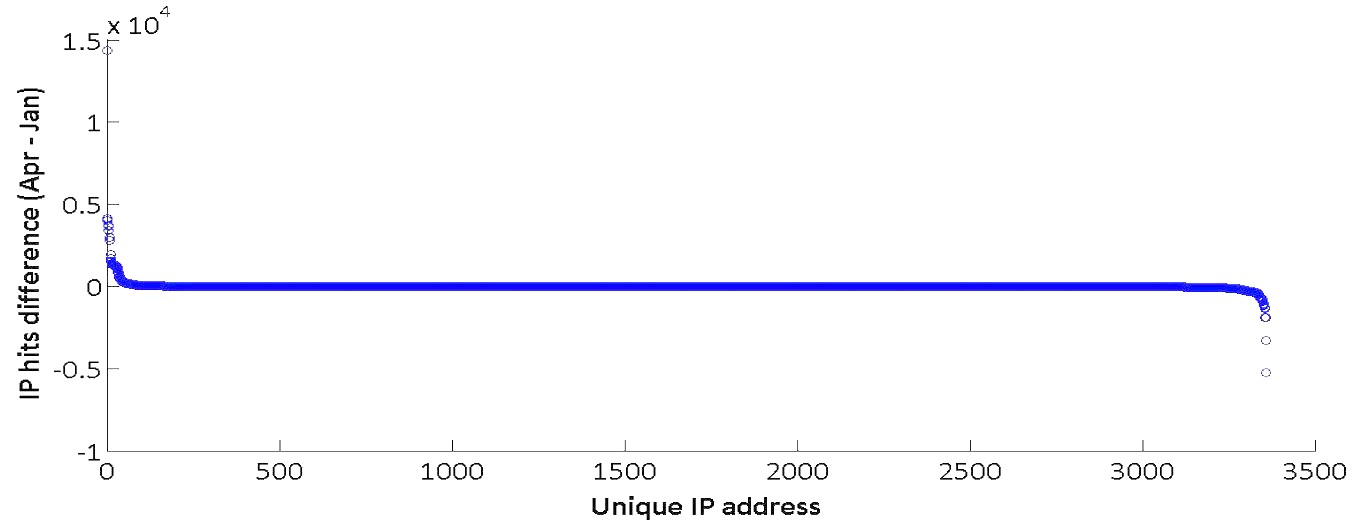}
\label{fig:learning1}
}
\hfil\hfil
\subfigure[]
{
\includegraphics[width=3.5in]{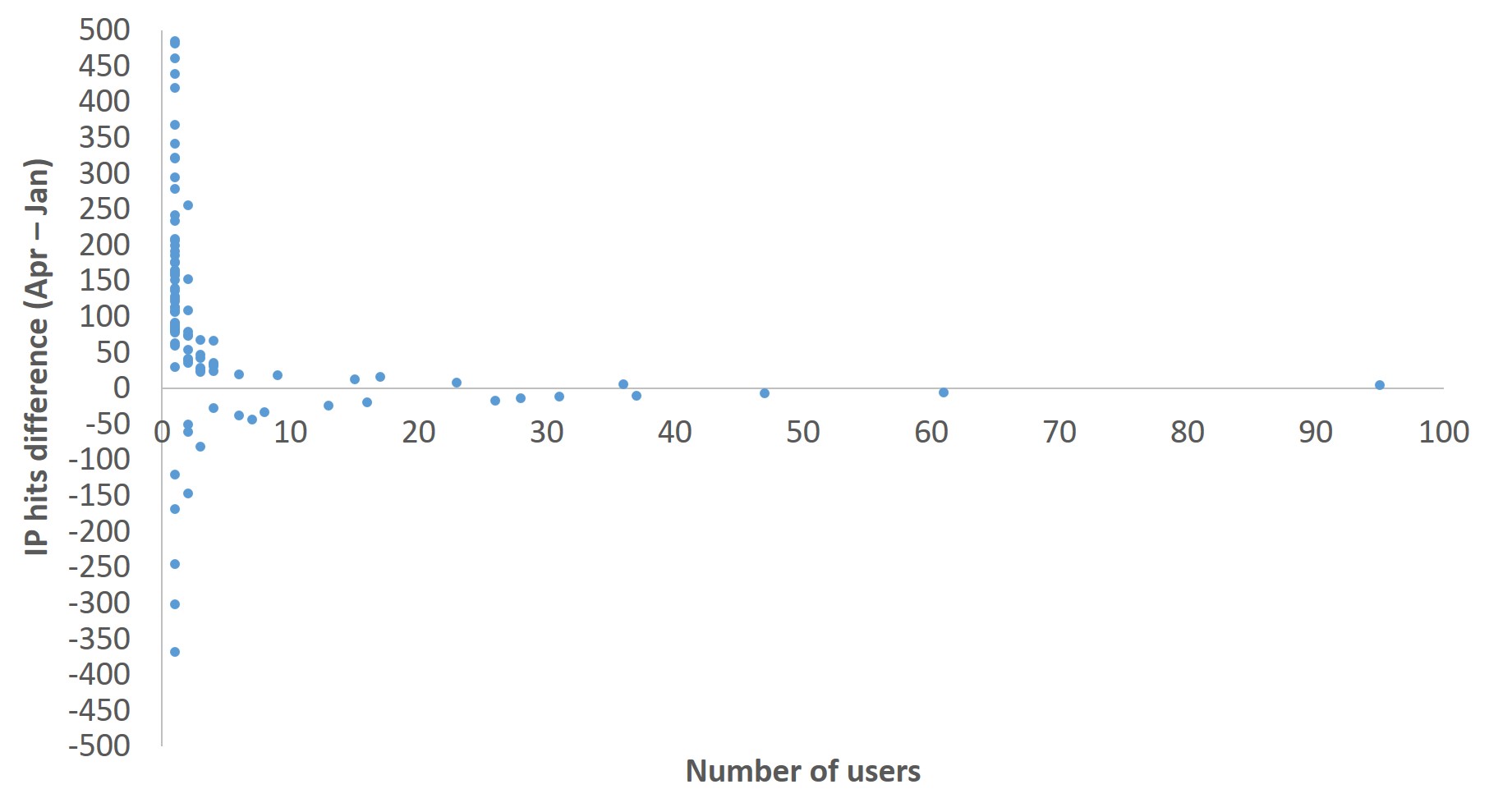}
\label{fig:learning2}
}
\caption{Scatter plot to show the difference in the URL hits of users in April and January 2014. (a) IP hits difference for 3,359 unique users dropped in April than in January 2014 which shows the success of landing page in training the users to avoid falling for phishing attacks; (b) Distribution of users whose difference in the number of clicks in January and April 2014 ranged from -500 to 500. We see large number of users with lesser clicks and few users with large number of clicks in April 2014.}
\label{fig:learning}
\end{figure*}

To measure the effectiveness of landing page in helping people not to fall for phishing attacks, we analysed the unique users (IP address) who were found in our dataset both in the months of January and April 2014 to observe their click stream. Figure \ref{fig:learning1} shows the learning curve for all such users where the x-axis represents 3,359 unique users and y-axis shows the difference in the number of URL hits a user made in April 2014 than in January 2014. We observed that 46\% users had lesser number of hits in April than in January 2014 which showed that the landing page was effective in helping users in guiding them not to fall for phishing attacks. With the difference in URL hits calculated as above, Figure \ref{fig:learning2} shows the distribution of users whose difference in number of URL hits ranged from -500 to 500. We saw that a large proportion of users clicked lesser number of URLs from January to April 2014, whereas less number of users clicked more URLs in April than in January 2014.

\subsection{Domain Analysis}
\begin{figure}[h]
\centering
\includegraphics[width=2.5in]{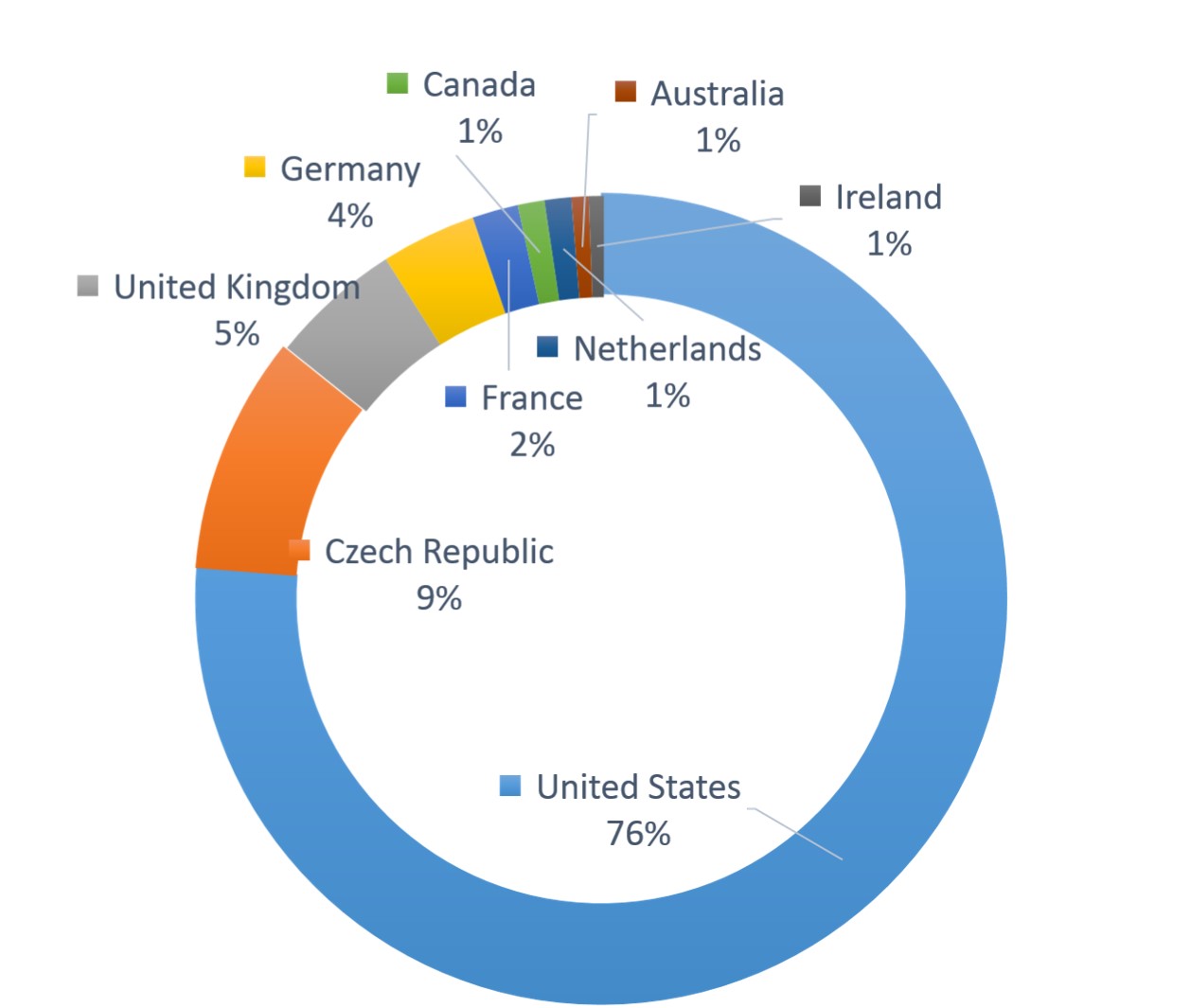}
\caption{Distribution of top nine countries hosting phishing domains in \textit{2014 dataset}. The United States is observed to host maximum number of phishing domains.}
\label{fig:countries}
\end{figure}

We parsed phishing URLs to get the corresponding domain from our datasets. We found 320 unique domains in \textit{2008 dataset} and 1,893 in \textit{2014 dataset}. We then performed WHOIS \footnote{http://en.wikipedia.org/wiki/Whois} lookups to collect information for each unique domain. We got information like the creation date, registrar name, and the information that a registrant enters while creating / registering a domain in his / her name like his / her name, location (street, city, state, country), email, and telephone. For each domain, we used Geolocation API, \footnote{http://ip-api.com/} to get access to the geographical location information of machines hosting the phishing domains. It gave information like IP address of the domain, location (latitude, longitude, city, region, country) where the domain is hosted, ISP information of the particular IP address / domain. Figure \ref{fig:countries} shows the distribution of the countries hosting the domains for \textit{2014 dataset}. We show only the top 9 countries hosting the domains, since others had less than 1\% share of hosting domains. The United States was observed to be the top-most country hosting the phishing domains. This is because a large percentage of the World's popular registrars / hosts like Godaddy.com, eNom, WildWestDomain.com etc. are hosted in the United States. \footnote{http://www.webhosting.info/webhosts/tophosts/global/} In \textit{2008 dataset}, countries like Hungary, France, Belarus, and Australia were found to host maximum number of phishing domains after the United States.

\begin{figure}[h]
\centering
\includegraphics[width=3.5in]{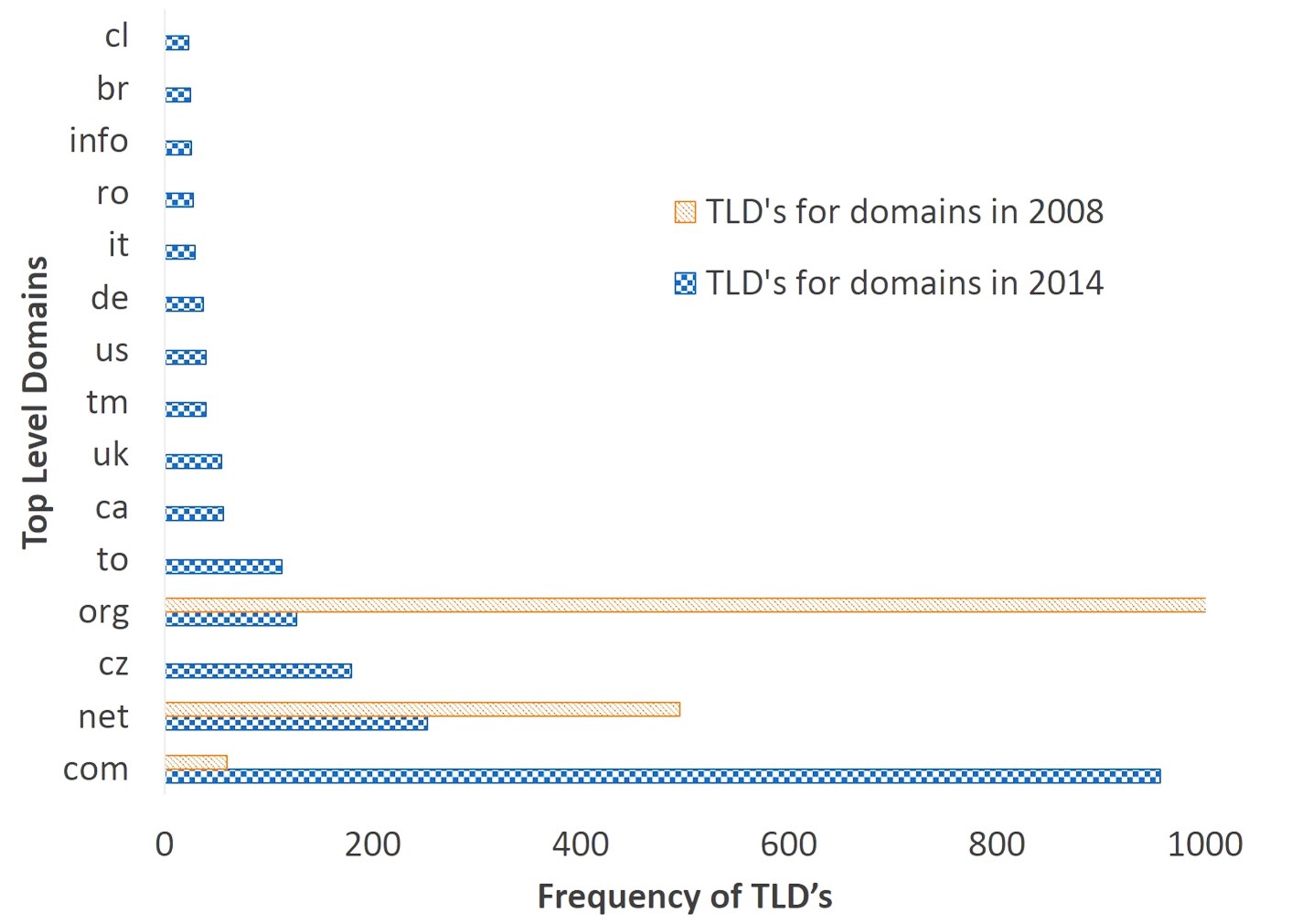}
\caption{Distribution of Top Level Domain (TLD) for phishing domains observed in \textit{2008} and \textit{2014 datasets} having frequency of occurrence greater than 20. `.org' is found to be the popular choice amongst phishers now to spread phishing.}
\label{fig:tld}
\end{figure}

We found 35\% of the phishing domains hosted on a single server (same IP address) maintained by FreeDNS, a service operational in the United States, \footnote{http://freedns.afraid.org/} which allows people to share their domain with others, and similarly use other's domains. It allows anyone to add multiple subdomains off a primary domain till the owner of that domain specifically disallows it. Such kind of ``subdomain registration services" offer users a ``domain name" -- their own DNS space -- and often offer free DNS management. As a result, it affects the innocent users as well whose domains can be exploited by the phishers in carrying out their attacks. APWG reported a similar service operational in Netherlands \cite{apwg}. Use of subdomain services continues to be a challenge for ISPs tracking down the phishing URLs, because many of these services are free, offer anonymous registration, and only the subdomain providers themselves can effectively mitigate the phishing attacks.
\newline\indent
We parsed the domains to get the Top Level Domain (TLD) corresponding to each unique domain. Primarily, the choice of TLD depends upon a number of factors such as the nature of the website, type of business, location of the business, targeted audience, sought after territory, type of organization, and the like. The distribution of TLDs observed in our dataset having frequency greater than 20 (arranged in increasing order) is shown in Figure \ref{fig:tld}. `.org' was the most popular TLD in 2008 since any person / entity was allowed to register for it. \footnote{http://en.wikipedia.org/wiki/List\_of\_Internet\_top-level\_domains}
With time, `.com' has become the main TLD for domain names, since many people believe that if net surfers do not remember the extension, they are more likely to type .com in front of the domain name. \footnote{http://www.pcnames.com/Articles/Common-TLDs-and-Their-Uses} We found TLDs like .au, .de, .us and various other country code TLDs (ccTLD) in our dataset, that are geographically specific and can be obtained at cheaper annual subscriptions compared to that of generic TLDs. However, their percentage remained low since phishers want to reach to larger population which could be achieved through popular domains.

\begin{figure*}[t]
\centering
\subfigure[]
{
\includegraphics[width=3.0in]{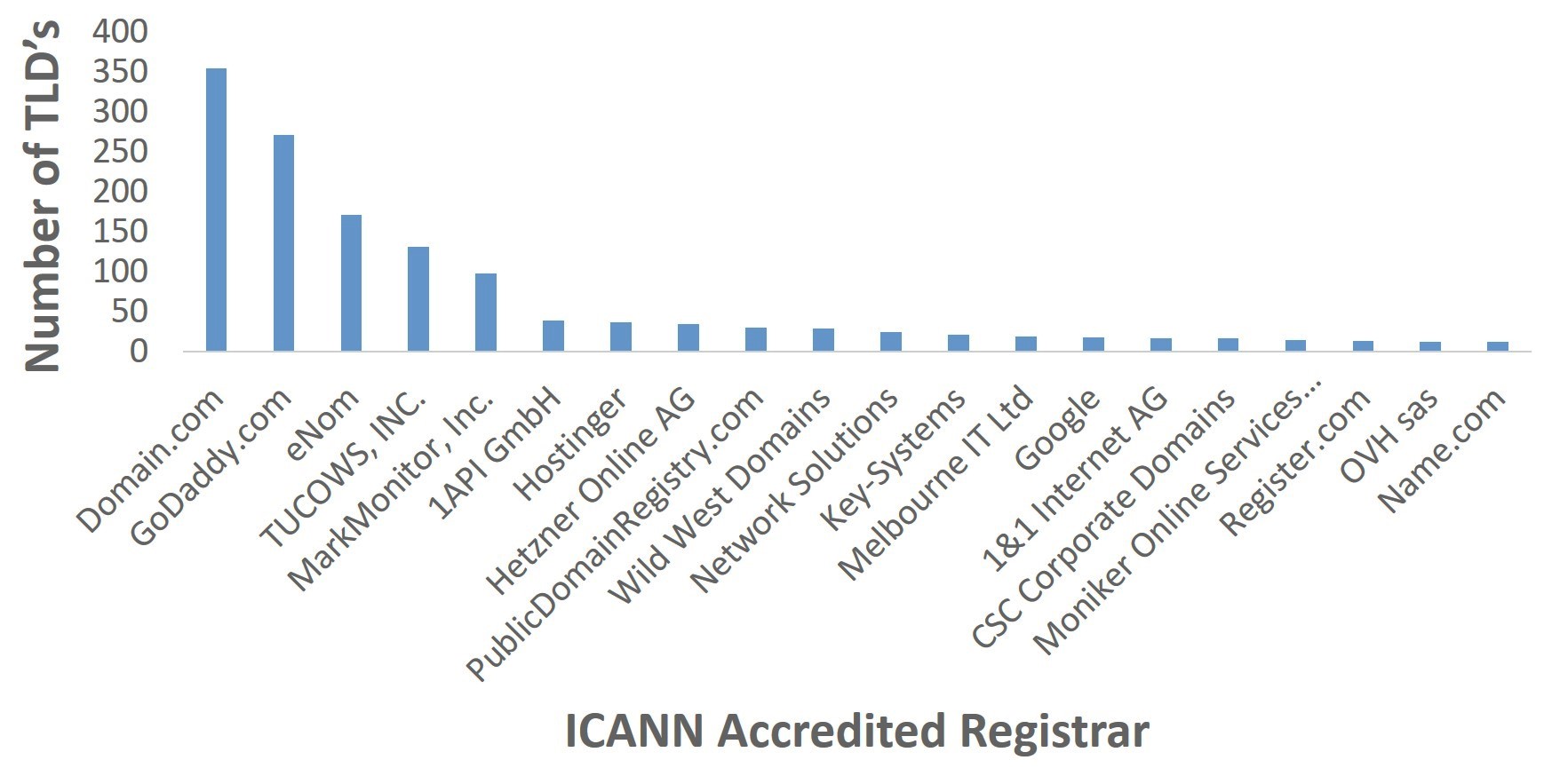}
\label{fig:icann}
}
\hfil\hfil
\subfigure[]
{
\includegraphics[width=3.0in]{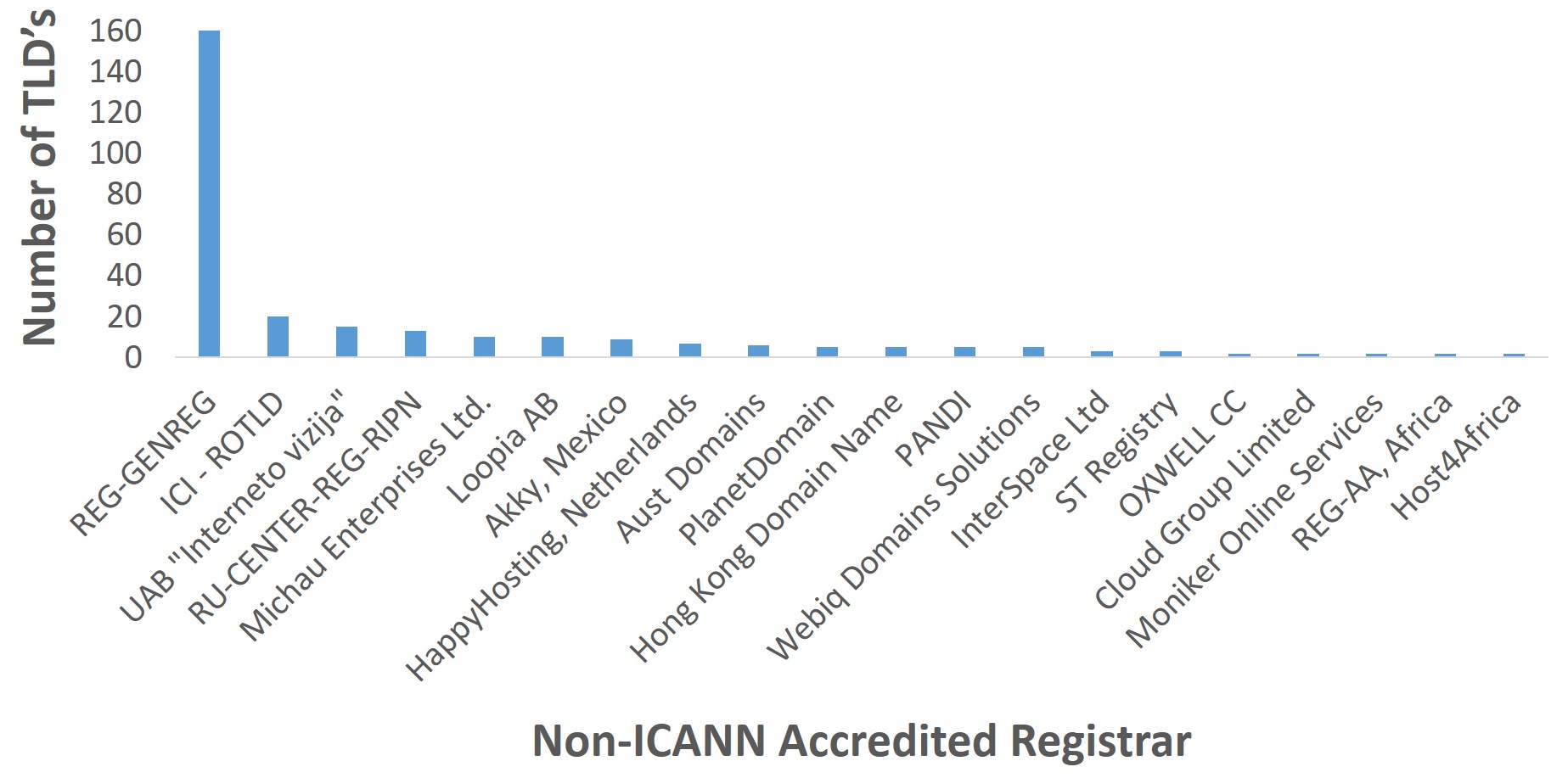}
\label{fig:non-icann}
}
\caption{Frequency distribution of registrars used by phishers to register their domains as observed in \textit{2014 dataset}. (a) Top 20 ICANN accredited registrar; (b) Top 20 non-ICANN accredited registrar.}
\label{registrar}
\end{figure*}

Domain names today can be registered through different companies, which compete with one another on the basis of price, value-added services and customer service. ICANN coordinates the Internet's global domain name system and maintains a list of accredited domain name registrars.~\footnote{http://www.internic.net/alpha.html} These registrars are associated with ICANN with certain rules and regulations, and applicable laws. We compared this list with the registrars recorded in our dataset and show the top 20 registrars in Figure \ref{fig:icann}. However, there are some registrars who do not register themselves under ICANN. This could be done to prevent their activities from being monitored and escape from the annual fee subscription charged by ICANN. The top 20 non-ICANN accredited registrars found in our dataset are shown in Figure \ref{fig:non-icann}. We found that 75.6\% registrars were accredited to ICANN which accounted for 85.6\% of the total URLs clicks in our \textit{2014 dataset}. About 55\% ICANN accredited registrars were found in \textit{2008 dataset}. The increase in use of ICANN accredited registrars by phishers shows that these accredited registrars are not properly monitoring the domains registered under them and are hence being exploited by the phishers to carry out their spamming activities. Phishers impersonate as a domain name registrar and try to get access to registrants's domain account credentials. \footnote{https://www.icann.org/en/system/files/files/sac-028-en.pdf} This shows the impending need of proper surveillance to keep criminals out of such activities.

\begin{figure}[h]
\centering
\includegraphics[width=3.5in]{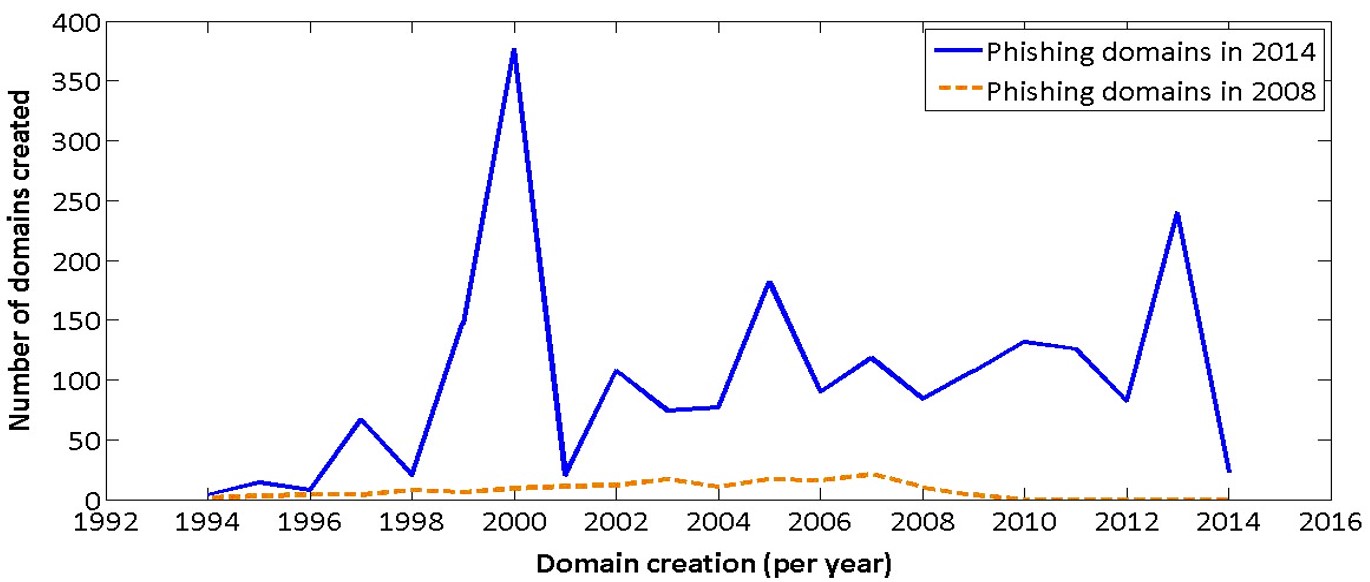}
\caption{Frequency distribution of phishing domains created per year for \textit{2008} and \textit{2014 datasets}. We see a jump in the number of domains created now than earlier to spread phishing for longer durations.}
\label{fig:domain_creation}
\end{figure}

Next, we saw phishing ``uptime" / ``live" time, i.e. the time until phishing attacks remained active \cite{apwg}. To analyse this, we plotted the domain creation dates, obtained from WHOIS lookup, as shown in Figure \ref{fig:domain_creation}. We observed that domains registered for the purpose of phishing became active almost immediately upon registration, as some of the phishing domains recorded in \textit{2014 dataset} were created in 2014. At the same time, we saw that ISPs have developed good techniques / learning to mitigate phishing attacks since they could detect the domains even when the window to track suspicious domain registrations from the perspective of phishing was very small. This shows that the phishing uptime has reduced. APWG reported that average phishing uptime has reduced to 8 hours in later half of 2013 \cite{apwg} from 20 hours as observed in 2008 \cite{apwg10}. It would have been possible for us to calculate the exact phishing time as observed in 2014, if we would have the information from ISPs about when the phishing website was observed first for attack. We also found that the proportion of domains created in 2014 is more than what was observed in \textit{2008 dataset}. It shows that phishers have started to create greater number of domains to keep their activity continuous for longer duration.

\subsection{Browser Analysis}
\begin{figure}[h]
\centering
\includegraphics[width=3.7in]{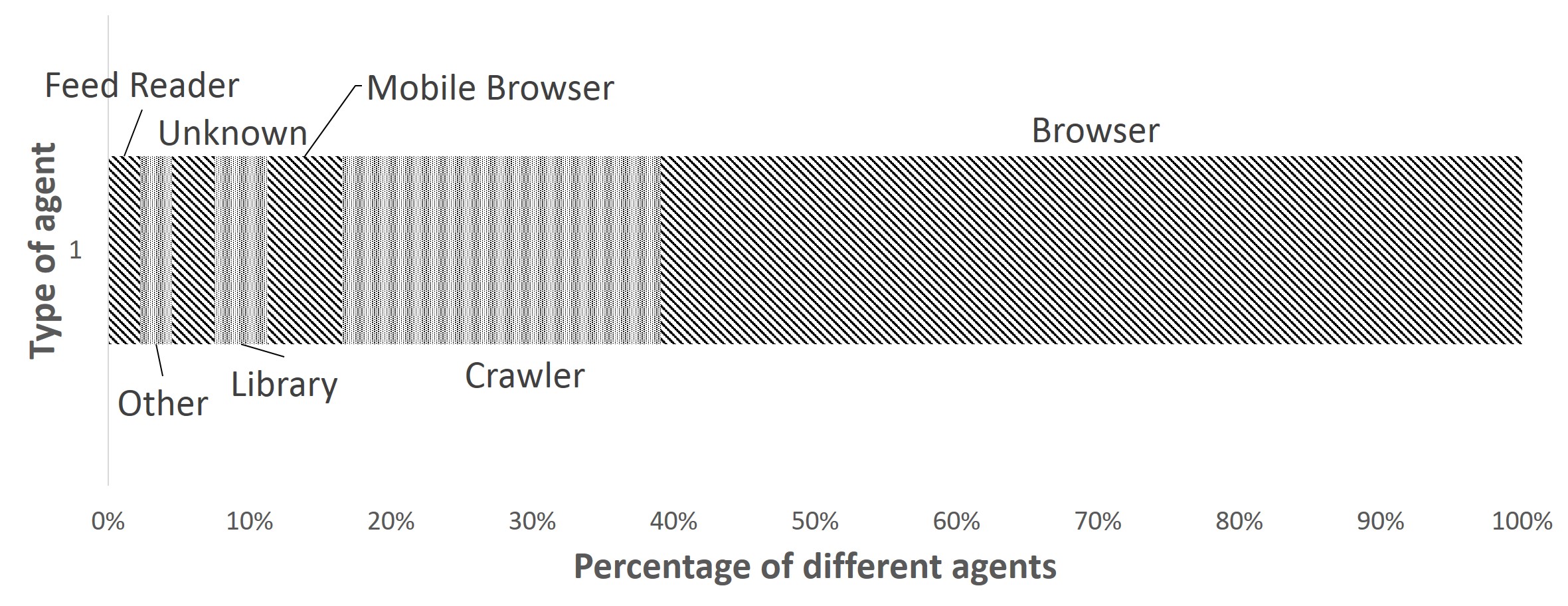}
\caption{Distribution of different types of user agents associated with the victims as observed in \textit{2014 dataset}. People are falling for phishing even while accessing their mobile phones.}
\label{fig:browser}
\end{figure}

We parsed the browser / client information logged in redirect logs to analyse the distribution of different user agents used by victims at the time of attack and to study the effectiveness of blacklists in controlling phishing attacks today. Web browsers have free add-ons (or ``plugins") that can help people detect phishing sites by giving warnings. We used User Agent Parser API, \footnote{http://www.useragentstring.com/pages/api.php} which gave information regarding the client (victim) software type (browser, crawler, web browser etc.), Operating System (OS) used by the client (victim). Figure \ref{fig:browser} shows the distribution of various type of agents used by the victims. We see that 60\% of user hits come from web browsers, even after having built-in plugins.

Table \ref{tab:user_agent} shows the information of top 5 browsers and top 5 crawlers having the highest frequency of occurrence recorded in our dataset. Though Internet Explorer and Mozilla Firefox have plugins to detect blacklisted URLs, these were the most used user agents when victims clicked the phishing URLs. Researchers have shown that browser blacklists are not efficient enough to detect the phishing URLs \cite{egelman} \cite{sheng}. We observed similar trends in our \textit{2014 dataset}.

\begin{table}[h]
\centering
\caption{Frequency distribution of top 5 browser agents and crawlers as observed in \textit{2014 dataset}. Internet Explorer and Firefox are observed to get maximum clicks even though they have blacklists installed. This shows that blacklists today are not very effective in detecting phishing links.} \label{tab:user_agent}
\begin{small}
\begin{tabular}{|p{2cm}|p{1.20cm}|p{1.75cm}|p{1.5cm}|} \hline
\small Browser (60\%) &\small Frequency &\small Crawler (23\%) &\small Frequency\\ \hline
Internet Explorer & 40\% & Bingbot & 25.9\% \\ \hline
Firefox & 19.8\% & Java & 21.4\% \\ \hline
Google Chrome & 14.3\% & Ezooms & 18.3\% \\ \hline
Safari & 7.2\% & Baidu Spider & 8.2\% \\ \hline
Opera & 3\%	& Googlebot & 7.5\% \\ \hline
\end{tabular}
\end{small}
\end{table}

While looking at various agents as obtained in the client / browser field from the redirect logs, we found suspicious activity with some of the IP addresses. They were identified as crawlers (Googlebot, BaiduSpider, MSN bot, Java crawler etc.) by the user agent string API that we used. These crawlers ran scripts (bots) to increase the number of redirection to the landing page. We identified 2,110 such unique IP addresses. Variants had consecutive numbers in either of the 3 subnets of the IP address with the first subnet being common. (For example, if one of the address IP address was 157.55.XXX.XXX, another one found was 157.54.XXX.XXX). There is a high possibility that they belonged to a common network. Majorly, these IP addresses were found from the United States, China, Japan, and Russia. Table \ref{tab:crawler} shows the information about top 5 IP addresses (\textit{in terms of their frequency}) that were recorded in the dataset.
\begin{table}[h]
\centering
\caption{Information regarding the crawlers as observed in \textit{2014 dataset}. We see that phishers used scripts to increase the number of redirections to the landing page.}
\label{tab:crawler}
\begin{small}
\begin{tabular}{|p{2.75cm}|p{1.5cm}|p{1.5cm}|} \hline

\small IP address format &\small Country &\small Agent type\\ \hline
157.55.XXX.XXX & Unites States & Bingbot \\ \hline
180.76.XXX.XXX & China & Baiduspider \\ \hline
199.30.XXX.XXX & United States & MSN bot  \\ \hline
123.125.XXX.XXX & China & Baiduspider  \\ \hline
176.195.XXX.XXX & Russia & Googlebot  \\ \hline

\end{tabular}
\end{small}
\end{table}
\newline\indent
Around 83\% of IP addresses which ran such scripts did not have any URL in their request header to the landing page. These IP addresses could be possibly some phishing kits that had linked content (like images and text) to their script from the original landing page and are hence redirected to the landing page. However, its difficult to draw a conclusive opinion until and unless we get some information about the IP address of phishing kits from the ISPs.

\subsection{Referrer Analysis}
To look at the breeding zones i.e., the actual places that were actively used by phishers to spread their phishing links, we analysed the referrer information available from the redirect logs. Referrer is the web page or application that contains the link to web page pointed by the phishing URL. We aggregated the referrer clicks by grouping all the unique referrers together and listed them in Table \ref{tab:referrer}.
\begin{table}[h]
\centering
\caption{Top 10 referrers of phishing URLs based on the number of clicks as observed in \textit{2014 dataset}. We see that phishers are targeting online social media to spread phishing URLs.}
\label{tab:referrer}
\begin{small}
\begin{tabular}{|p{5cm}|p{1.15cm}|} \hline
\small Referrer &\small Clicks \\ \hline
http://www.google.com & 980\\ \hline
http://m.facebook.com & 670\\ \hline
http://fasebook.c0m.at & 640 \\ \hline
http://www.facebook.com & 550\\ \hline
http://www.clixsense.com & 220 \\ \hline
http://www.youtube.com & 181 \\ \hline
http://servinox.com.co & 132 \\ \hline
http://www.akihabarashop.jp & 130 \\ \hline
http://dflogins.1s.fr & 91 \\ \hline
http://www.google.ro & 90 \\ \hline
\end{tabular}
\end{small}
\end{table}
\newline\indent
As observed in \textit{2008 dataset}, phishers used various advertising and blogging websites to spread phishing URLs. Websites like \textit{http://www.12gbfree.com/}, \textit{http://cache.phazedll.com}, where users could post / search advertisements; blogging site like \textit{http://tipp-pez.blogspot.com/} were found to receive maximum clicks in the dataset. From the table, however, we see that phishers have shifted their target to online social media like Facebook and YouTube to propagate their phishing links. Similar results were obtained by Chhabra et al. in 2011 \cite{chhabra}. We also observed search engines like \textit{google.com}, \textit{yandex.com}, and \textit{bing.com} in our dataset which received high referrer clicks. Several country specific versions of Google like \textit{google.com.au}, \textit{google.com.uk}, \textit{google.com.de} etc. were also recorded in the dataset. This might be possible because users tried to search more information about the phishing URLs on these search engines and hence were redirected to the landing page from there. The referrer having the third highest click, \textit{http://fasebook.c0m.at}, was built on an illegal phishing domain, c0m.at, registered in France. We could not collect much information on it since it was de-activated by the hosting service. \footnote{http://whois.venez.fr/whois.fr.html?name=fasebook\&domain=c0m.at}
We also observed landing page as referrer string in the dataset, \textit{http://phish-education.apwg.org/} and \textit{education.apwg.org} without containing any URL in the requesting header. These referrer could be present because of two reasons, (i) if the user clicked on some image available on the landing page and hence got redirected to the same page, or (ii) the phisher might have linked the text / image from the landing page to its fake website. In this case, when the user's web browser is loaded, the images are loaded from the actual server and the fake website is logged as a referrer in the original server's access log records. \footnote{http://www.symantec.com/connect/blogs/quotrefererquot-field-used-battle-against-online-fraud} However, since these referrers did not contain any URL (phishing), they were not considered in our analysis.

\subsection{E-mail feature Analysis}
\begin{figure}[h]
\centering
\includegraphics[width=3.5in]{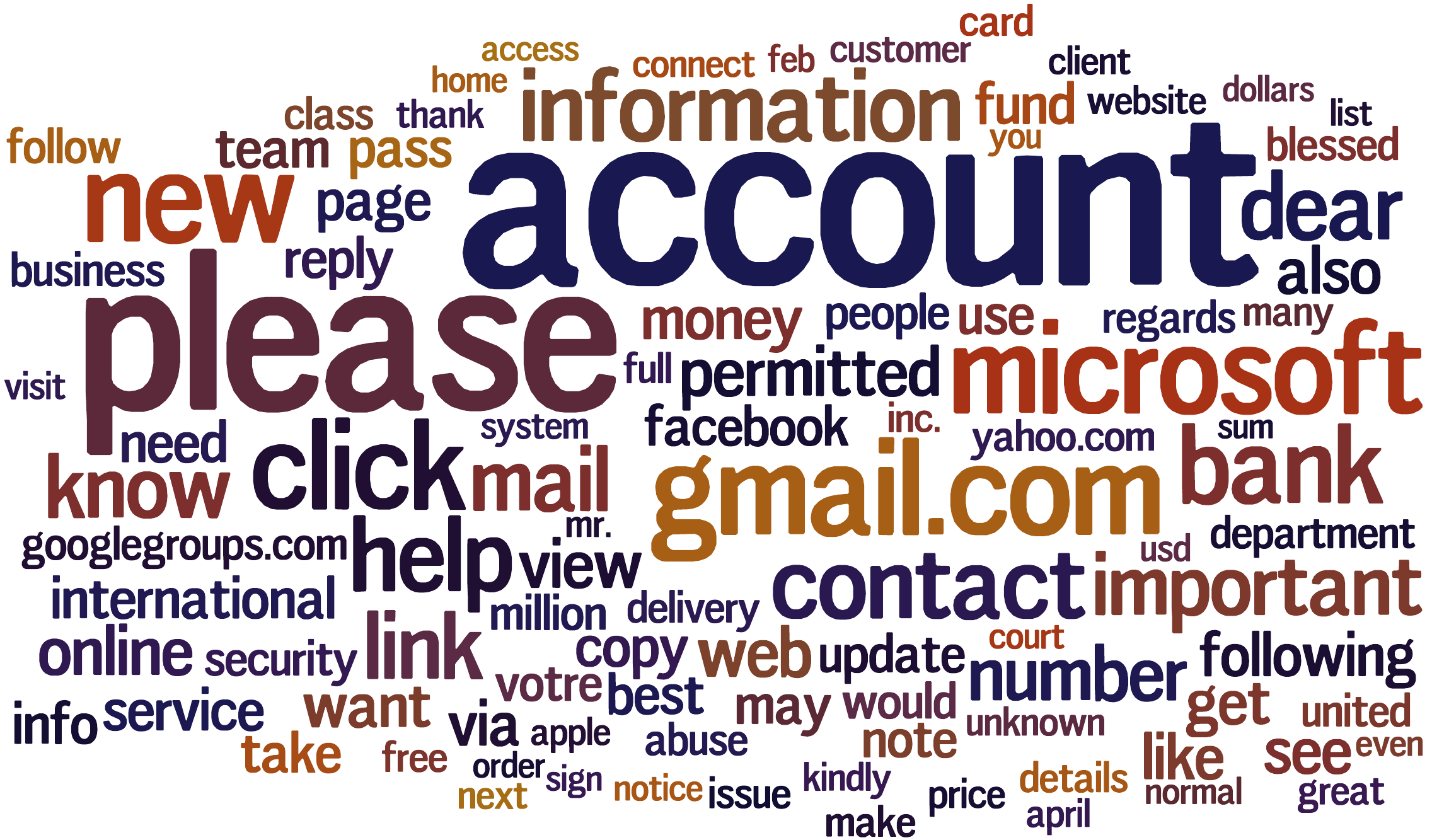}
\caption{Tag cloud of top 100 frequently occurring words as observed in the body of phishing e-mails archive for January - April 2014.}
\label{fig:tag_cloud}
\end{figure}

We received phishing e-mails from APWG where people / organizations / vendors reported the phishing e-mail to them.~\footnote{http://www.apwg.org/report-phishing/} To study features of the e-mail that contained URLs being redirected to the landing page, we compared set of domains corresponding to URLs in the e-mail feed, and in our redirect logs. We searched the APWG e-mail feed for e-mails containing the URLs, parsed them to get corresponding domains and fetched the e-mails containing the common domains. This was done to analyse features in the e-mails that led victims to click the phishing URL. We found 170 matches in our log data and in the APWG feed for \textit{2014 dataset}. We manually examined these e-mails and analysed the features in these e-mails. Some e-mails were present in different language. We used Google translate \footnote{https://translate.google.co.in/} to decode these e-mails. To study the change in techniques employed by phishers to send phishing e-mails, we compared the features obtained in our 2014 phishing e-mails dataset to those reported by Kumaraguru et al., who analysed the e-mail features in 2008 from the redirect logs of the same landing page \cite{pk-404}.
\newline\indent
Most of the e-mails had logos and banners to look more legitimate. As Dhamija et al showed \cite{dhamija}, the fact that these logos and banners look legitimate is the main reason people fall for phishing e-mails. We found that phishing scammers are using e-mail address to promote their fraud and scam. Around 25\% e-mails were found to target popular organizations like Microsoft, Paypal, Instagram etc. Phishers pretended to belong to these organizations and told users that their accounts were at risk and they need to upgrade / verify it. The subject line often had urgent and compelling words like ``Upgrade your account", ``Your online Banking Account is Locked - ACT NOW!", ``Suspicious activity found - IMPORTANT". The body of the e-mail had the link to upgrade the account where they were asked to give their e-mail and password. Figure \ref{fig:tag_cloud} represents the tag cloud of top 100 frequently occurring words found in the body of phishing e-mail dataset. We found the dominance of keywords like \textit{account}, \textit{information}, \textit{bank}, \textit{microsoft}, \textit{update} etc. This shows the presence of financial fraud in phishing e-mails.
Some of the messages also had malicious attachment which the user had to download in order to update their details. All the e-mails mentioned that the account would be locked or shut down if the recipient failed to act within the given time frame. Around 80\% e-mails gave a deadline of 24 - 48 hours for the user to act on. One common message that was observed in the message regarding the timeline was ``Your account has been compromised. Please update it else it will be closed within 24 hours". Kumaraguru et al. reported majority of such e-mails in 2008 \cite{pk-404}.
\newline
\indent
Phishers have brought change in their techniques to lure more users in their trap as compared to 2008. Around 26\% e-mails were used as promotional e-mails where phishers showed some attractive offers, deals of the day, business proposals, Facebook page recommendations, YouTube links etc. to make people fall prey to their attacks. They tried to motivate people by saying that they could get heavy discounts if they were first to click on the link. These messages often contained several links where the users were required to click to get more information. These were actually pointing to phishing links. Strong subject lines like ``UNBELIEVABLE – Check inside", ``Interesting Stuff" were used to convince people to click the suspicious links.
Some of the e-mails tried to lure victims by giving them monetary benefits. We observed two kind of e-mails related to financial relief, (i) 11\% of the e-mails tried to fool users by telling them that their e-mail was chosen (random) to win a cash prize of X amount. The prize quoted was high enough to sway them to open and respond to the e-mail. We observed keywords like \textit{million}, \textit{dollar}, \textit{usd}, \textit{fund} etc. in the tag cloud for top 100 words in phishing e-mails (Figure \ref{fig:tag_cloud}). Users were asked to click a link to redeem the offer which was actually a phishing link, (ii) Another kind of e-mail (9\%) that we observed was, phishers tried to play with the emotional sentiments of people to convince them to click their fraudulent links. The subject line was often read as ``Please help me", where phishers tried to build stories like ``My X relative died, I want to transfer my money to your country. Please help me by giving your bank details". Here, phishers tried to portray as really helpless and urged people to respond and help.
\newline \indent
We also found around 8\% e-mails which were random requests to add people on network. People sent their pictures, links (suspicious), videos asking users to know more about them and connect to them. These links were actually linking to the phishing sites. We found that majority (75\%) of such e-mails contained derogatory words and grammatical errors in the message content. Our phishing e-mail archive feed for 2014 also recorded 6\% e-mails that contained message like ``Your order is shipped. Click here for more details". The e-mail body had the details of the product and users were asked to confirm their order by clicking the link given in the e-mail. Phishers targeted popular online shopping sites like Amazon, ebay etc.
\newline \indent
This shows that phishers are trying to target users emotionally and financially to make their phishing campaigns successful. However, some of the strategies to con people to fall for phishing remained the same as observed in 2008, like pretending as a popular financial institution and compelling them to click the URL given in the e-mail to keep their accounts updated.

\subsection{Observations}
In this section, we discuss some of the observations that we obtained from \textit{2014 dataset}. As shortened URLs are extensively used in online social media to carry out phishing attacks \cite{chhabra}, we looked for the presence of shortened URLs in our \textit{2014 dataset}. APWG reported an increase in use of URL shortening services in later half of 2013 from the first half. The list of URL shortening services found in our \textit{2014 dataset} as shown in Table \ref{tab:url-shortener}, however, shows a decline in the use of these services by phishers. We used Bitly API \footnote{http://dev.bitly.com/links.html\#v3\_shorten} to check if the long URL was shortened by Bitly or not. The API gives a hash value corresponding to the URL. It returns a value `0' if the URL is already shortened and `1' if it is shortened for the first time. We queried all unique URLs in the dataset to find their hash values. For the rest of the services, we checked if the URL contained the string (shortening service) or not. This is because these services do not provide an API to check if the URL is already shortened or not.
\begin{table}[h]
\centering
\caption{URL shortening services as observed in \textit{2014 dataset}. We observed a decline in use of URL shortening services.}
\label{tab:url-shortener}
\begin{small}
\begin{tabular}{|p{2cm}|p{1.5cm}|} \hline
\small URL shortening services &\small \% of URLs shortened \\ \hline
Bit.do & 2\%\\ \hline
Bit.ly & 1.1\%\\ \hline
Goo.gl & 0.7\%\\ \hline
Tiny.cc & 0.03\% \\ \hline
Youtu.be & 0.02\% \\ \hline
short.ie & 0.01\% \\ \hline
\end{tabular}
\end{small}
\end{table}
\newline
The decrease in the use of URL shortening services could be because people might have lost faith in these URL shortening services and hence phishers are discontinuing to use them.
\newline\indent
We analysed number of domains which are malicious in our \textit{2014 dataset}. For this, we looked at the domains marked malicious by VirusTotal and SURBL. VirusTotal is a free online service that characterize a URL / domain as malicious using 52 different website / domain scanning engines and datasets.\footnote{https://www.virustotal.com/en/about/credits/} SURBL is an aggregated list of websites that have appeared in unsolicited messages. \footnote{http://www.surbl.org/} Both these services provide an API to check whether the URL / domain is malicious or not. We queried all the unique domains against these two services and found 12.5\% domains were registered malicious on VirusTotal, and 3\% on SURBL. This shows that these services are not prudent enough to detect malicious / bad domains.

\section{Conclusion}
Our preliminary analysis shows the change in techniques incorporated by phishers to launch attacks. We saw that phishers are improving their techniques by making their URLs look more genuine and legitimate to convince people to click on them. They prepend authentic-sounding words with the domains in the URL to convince people that it's genuine. Phishers have also increased the numbers of phishing domains registered per year to keep propagating their activities for longer duration. We observed that majority of registrars were accredited under ICANN which is bound under proper rules and regulations. However, the ability of phishers to attack them shows the impending need of proper surveillance to decrease criminal activities. We found that many country specific domains (low cost) are created which aim to target local population. But, the number of such domains remain low since phishers want to capture larger proportion of people by using international domains.
\newline\indent
We saw that browser blacklists today are not able to detect the phishing links. Internet Explorer and Firefox, though have several built-in plugins to detect phishing sites, we found large number of such user agents when victims clicked the phishing URLs. We observed that phishers have shifted their focus from advertising and blogging websites and are now targeting social media to spread their phishing links. We also saw change in phishing e-mails used to con people in clicking phishing URLs. Phishers have come up with some new techniques like sending promotional and monetary-related e-mails to lure victims to give out their personal information. Besides, some phishing e-mails have not changed considerably with time, largely asking victims to click on links (phishing) to upgrade their account. We also observed the effectiveness of landing page in helping people to avoid falling for phishing attacks. Forty six percent users clicked lesser number of phishing URLs in April than in January 2014, which shows the landing page was successful in helping and guiding them not to fall for phishing attacks.
\newline\indent
Our results will help ISPs and other financial / government organizations to enhance their techniques to detect phishing URLs. As we found that phishers are evolving their techniques to construct a phishing URL, take down vendors can bring change in their methodology to mark a website as phishing and take it down. ICANN accredited registrars can build stricter policies to mitigate the unauthorized use of their services by phishers. Our results showed that users are learning from the landing page, we believe, it will benefit a larger population if more ISPs adopt this initiative and redirect their users to the landing page.
\newline\indent
For this paper, we only had access to logs and e-mails, so we don't have any information available about the users arriving at the landing page. It would be interesting to study the behavioral patterns of these users to know what preempted them to click these links. It would have been possible for us to do a detailed analysis on phishing kits used in 2014, if we would have more information about them like IP address, type etc. Since the phishing sites were already being redirected to the landing page, we could not analyse the contents of these websites to see in what respect they looked different from the legitimate sites. It restricted us to analyse the features and patterns of these phishing sites that prompted ISPs and take down vendors to mark a website as phishing, and hence bring it down. It would be interesting to do a longitudinal study on the dataset available from 2008 (when the landing page was deployed) to 2014, to reveal emerging phishing patterns. Even though several classifiers are built to classify phishing and non-phishing URLs, people still fall for phishing attacks. We plan to extend this analysis to generate features that could help in building a plugin / service to predict phishing and non-phishing URLs on the fly, and help users not to click phishing URLs.

\section*{Acknowledgment}

The authors would like to Peter Cassidy and Guhan Iyer from APWG to provide access to the data. We would like to express our sincerest thanks to all members of Precog, Cybersecurity Education and Research Centre (CERC) at IIIT-Delhi, for their continued support and feedback.



%
\bibliographystyle{abbrv}
\bibliography{bibfile}

\end{document}